\documentclass[10pt]{revtex4}
\usepackage{hyperref}

\begin{document}

\title{\Large Equivalence between two different field-dependent BRST
formulations }

\author{ Sudhaker Upadhyay}
\email{ sudhakerupadhyay@gmail.com}
\affiliation { Department of Physics, Indian Institute of Technology Kanpur, Kanpur 208016, India}
\author{ Bhabani Prasad Mandal}
\email {  bhabani.mandal@gmail.com}

\affiliation { Department of Physics, 
Banaras Hindu University, 
Varanasi 221005, India   }

\begin{abstract}
Finite field-dependent BRST  (FFBRST) transformations were constructed by integrating infinitesimal
  BRST transformation in a closed form. Such a generalized transformations have been extended in various branch of physics and found many applications. Recently BRST transformation 
  has also been generalized with same goal and motivation in slightly different manner.
  In this work we have shown that the later formulation is conceptually equivalent
  to the earlier formulation. We justify our claim by producing the same result 
  of later formulation using earlier FFBRST formulation.
 
\end{abstract}

\maketitle 
\section{Introduction}
The  BRST  symmetry transformation is a fundamental tool for the study of 
gauge theories \cite{brst, tyu}. This symmetry guarantees the quantization, renormalizability, unitarity and 
other aspects of Yang-Mills (YM) 
theories \cite{ht,wei,masud}. 
The derivation
of Slavnov-Taylor identities \cite{1,2} in YM theory utilize the  BRST symmetry 
transformation. 
In Hamiltonian formulation, the invariance of the theory under the gauge transformation of variables generated by the
constraints is replaced by the invariance under the
transformations generated by the Batalin-Fradkin-Vilkovisky (BFV)-BRST charge \cite{ht}.
 Action and measure of the
related path integrals should be invariant under the BFV-BRST transformations.

The  concept of FFBRST  transformation was first introduced by
Joglekar and Mandal in 1995 \cite{bm} in which they   explicitly   show that the
 Jacobian  of path integral measure could be written exponential of some local
 functional of fields. This local functional of fields can be adjusted to a desirable 
 form by constructing the finite field-dependent parameter appropriately. Due to this interesting result   the FFBRST transformation is capable to relate the generating
 functionals of different effective theories. After this seminal work the FFBRST transformation
have found great applicability in diverse area of gauge theories  \cite{mru,sdj1,rb,susk,jog,sb1,sb11,sud10,sud11,sud12,smm,sudd,sud,fs0,fs,rbs,rbs1,skbm}. 
 For instance,  a correct prescription for   poles in the gauge field propagators in noncovariant gauges has been derived by connecting  covariant  and noncovariant gauges for
 YM theory through FFBRST transformation \cite{jog}.
 The long outstanding problem of divergent energy integrals in Coulomb gauge was also settled
 by FFBRST transformation \cite{sdj1}.
The YM theory in the non-perturbative infrared region is plagued by the Gribov ambiguity \cite{gri}. This ambiguity has been fixed by proposing the  Gribov-Zwanziger theory \cite{zwan}  which restricts the domain of integration to the boundary called as Gribov
first horizon. The FFBRST transformation has also  played  a crucial role in connecting the YM theory to Gribov-Zwanziger theory \cite{sb, sb2}.  
The  connections of the linear and the  non-linear gauges for the  perturbative theory of quantum gravity \cite{sud12} and for Bagger-Lambert-Gustavsson (BLG) theory \cite{fs}   
 at both classical and quantum level have been studied recently via FFBRST technique.
Furthermore,  
 the gauge-fixing and ghost terms corresponding to
 Landau as well as maximal Abelian gauge
 for the  Cho--Faddeev--Niemi (CFN) decomposed theory \cite{cho,du,faddeev}, which enables us to explain and understand some of the low-energy phenomena by separating the   topological defects in a gauge-invariant manner,
appear naturally within  FFBRST formulation \cite{sud10}.
Moreover, the gaugeon formulation has also been justified through FFBRST transformations \cite{sb11,sud11}.

 Recently, Lavrov and Lechtenfeld have worked on field-dependent BRST transformation  \cite{lav}
 exactly similar to FFBRST formulation of Ref. \cite{bm} in slightly different way to claim 
 it as a new formulation. The purpose of this work is to point out that their 
 formulation is not a new one but the something similar to one which has been done 20 years ago in \cite{bm}.
 In this paper we show the equivalence between   the original FFBRST formulation 
 developed by Joglekar and Mandal,  and the field-dependent BRST 
 formulation studied by Lavrov and Lechtenfeld. The field-dependent parameter in Ref. \cite{bm}
 is finite as the Green function of such quantity is calculated between
 a vacuum and a state with gauge and ghost field which is finite rather than infinitesimal.
  Jacobian  for the path integral measure has been calculated explicitly in terms of 
  the finite parameter of the BRST transformation  \cite{lav}.
We found that although the ways of calculating Jacobian are different
 these two approaches are producing same results. We
 justify our claim by producing the results of \cite{lav} through the FFBRST formulation.
 We show the connection between  different $R_\xi$ gauges using FFBRST transformation.
  
The  paper is organized  as follows. We provide a brief discussion on FFBRST formulation
in Sec. II.
Then, in section III,  we compare two different approaches of generalizing 
the BRST transformation.  In the last section, we draw conclusion. 

\section{FFBRST transformation}
In this section let us revise the celebrated FFBRST formulation
which has been found the enormous applications in gauge theories. 
To achieve this goal, we first define 
the usual BRST transformation for a generic field $\phi$ written collectively as follows:
\begin{equation}
\delta_b  \phi =s_b\phi\ \delta\lambda ={\cal R}[\phi] \delta\lambda,
\end{equation}
  where ${\cal R}[\phi]=s_b \phi$ denotes the Slavnov variation of $\phi$
  and $\delta\lambda$ is an anticommuting global parameter of transformation. 
Now, we write the  infinitesimal field-dependent transformation by making all the 
fields $\kappa$-dependent, where $\kappa (0\leq \kappa \leq 1)$ is a continuous parameter, as follows
\begin{equation}
\frac{d\phi(x,\kappa)}{d\kappa}={\cal R} [\phi (x,\kappa ) ]\Theta^\prime [\phi (x,\kappa ) ],
\label{diff}
\end{equation}
where the $\Theta^\prime [\phi (x,\kappa )]$ is an infinitesimal  field-dependent parameter.
Integrating the above infinitesimal field-dependent transformation from $\kappa =0$ to $\kappa= 1$, we obtain the following FFBRST transformation ($\delta_f$):
\begin{equation}
\delta_f \phi(x)\equiv \phi (x,\kappa =1)-\phi(x,\kappa=0)={\cal R}[\phi(x) ]\Theta[\phi(x) ],
\end{equation}
where 
\begin{equation}
\Theta [\phi] = \Theta ^\prime [\phi] \frac{ \exp f[\phi]
-1}{f[\phi]},
\label{80}
\end{equation}
 is the finite field-dependent parameter and $f[\phi]$ is given 
 by 
 \begin{eqnarray}
 f[\phi]= \sum_i \int d^4x \frac{ \delta \Theta ^\prime [\phi]}{\delta
\phi_i(x)} s_b \phi_i(x).
 \end{eqnarray} 
 
 Now, it has been found that the  FFBRST transformation having such field-dependent
 parameter leads  a non-trivial Jacobian for the functional measure of
  the generating functional   \cite{bm}.
We provide a glimpse of computing the Jacobian  of the path integral measure $({\cal 
D}\phi)$ in the functional 
integral under FFBRST transformation. In this regard, the first step is to 
define the functional measure under the FFBRST transformation as follows \cite{bm}
\begin{eqnarray}
{\cal D}\phi^\prime &=&J(
\kappa) {\cal D}\phi(\kappa).\label{jacob}
\end{eqnarray}
It has been shown explicitly in \cite{bm} that within functional integral the Jacobian $J(\kappa )$ can be replaced as 
\begin{equation}
J(\kappa )\longmapsto e^{{iS_1 [\phi(x,\kappa) ]}},\label{js}
\end{equation}
where $ S_1[\Phi ]$ is some local functional of fields, if and only if
 the  condition 
 \begin{eqnarray}
 \int {\cal{D}}\phi (x) \;  \left [\frac{d}{d\kappa}\ln J(\kappa)-i\frac
{dS_1[\phi (x,\kappa )]}{d\kappa}\right ] \exp{[i(S_{eff}+S_1)]}=0 \label{mcond}
\end{eqnarray} 
is satisfied. 
However, the local functional $ S_1[\Phi ]$
satisfies the following
initial boundary condition
\begin{equation}
S_1[\Phi]_{\kappa=0} =0.\label{ini}
\end{equation}
The infinitesimal change in Jacobian written in (\ref{mcond}) has the following explicit 
expression \cite{bm}
\begin{equation}
 \frac{d}{d\kappa}\ln J(\kappa)=-\int  d^4y\left [\pm\sum_i {\cal R}[\phi^i(y )]\frac{
\partial\Theta^\prime [\phi (y,\kappa )]}{\partial\phi^i (y,\kappa )}\right],\label{jac}
\end{equation}
where  we employ the positive ($+$) sign  for bosonic fields and however for fermionic fields 
we use the negative ($-$) sign.
  Hence the
  vacuum functional for a most general gauge theory defined by    
  \begin{eqnarray}
Z =\int {\cal D}\phi\ e^{iS_{eff}},
\end{eqnarray}
changes under FFBRST transformation due to Jacobian as
 \begin{eqnarray}
Z \left(\equiv\int {\cal D}\phi\ e^{iS_{eff}}\right)\stackrel{FFBRST}{---\longrightarrow}
Z' \left(\equiv\int {\cal D}\phi\ e^{i\left(S_{eff}+S_1\right)}\right),
\end{eqnarray}
where $S_{eff}$ is the effective action for this most general theory.
By constructing an appropriate parameter $\Theta$,  the expression for the functional $S_1$,
which extends the effective action, can be manipulated according to the requirement of the theory.
 
 \section{Comparison between different field-dependent BRST formulations: The YM theory}
Let us start the section by defining the vacuum functional for YM theory in $R_\xi$ gauge using the standard
Faddeev-Popov method as follows
\begin{eqnarray}
Z_\xi =\int {\cal D}\phi\ e^{iS_{\xi}},
\end{eqnarray}
where ${\cal D}\phi$ is the generic path integral measure written
 in terms of all fields involved in the theory  and the effective action, $S_{\xi}$, comprised with classical and gauge-fixed parts  is defined by
\begin{equation}
S_{\xi} = \int d^4x \left [-\frac{1}{4}F^{a\mu\nu}F_{\mu\nu}^a +\frac{\xi
}{2}(B^a) ^2 + B^{a}
\partial^\mu {A^a_\mu }+ \bar{C}^a\partial^\mu
D^{ab}_\mu C^b\right ]. 
\label{sf}
\end{equation}
 The covariant derivative is expressed as follows: $ D^{ab}_\mu \equiv \delta ^{ab}\partial
_\mu + g f^{abc}A^c_\mu $, where $g$ denotes the coupling constant. 

The usual BRST transformations under which the effective action (\ref{sf}) remains invariant are given by
\begin{eqnarray}
\delta_b A^{a}_\mu &= &  D_\mu^{ab}C^b\ \delta\lambda,\nonumber \\
\delta_b C^a &=&   -\frac{g}{2}f^{abc}C^b C^c\ \delta\lambda, \nonumber \\
\delta_b \bar{C}^a &=&  B^a \ \delta\lambda,\nonumber \\
\delta_b B^a &= &0.
\label{brst}
\end{eqnarray}
Using the above BRST transformations the effective action can also be expressed in terms 
of gauge-fixing fermion ($\psi$) as follows:
\begin{equation}
S_{\xi} = \int d^4x \left [-\frac{1}{4}F^{a\mu\nu}F_{\mu\nu}^a +s_b\psi\right ], 
\label{sf1}
\end{equation}
where the gauge-fixing fermion has following expression:
\begin{equation}
\psi =\bar{C}^a\left(\partial^\mu A_\mu^a +\frac{\xi}{2}B^a\right).
\end{equation}
Following the procedure outlined in Sec. III, the FFBRST transformations for YM theory are constructed by
\begin{eqnarray}
\delta_f A^{a}_\mu &= &  D_\mu^{ab}C^b \  \Theta[\phi],\nonumber \\
\delta_f C^a &=&   -\frac{g}{2}f^{abc}C^b C^c\ \Theta[\phi], \nonumber \\
\delta_f \bar{C}^a &=&  B^a \ \Theta[\phi],\nonumber \\
\delta_f B^a &= &0,
\label{ffb}
\end{eqnarray}
where  $\Theta[\phi]$ is an arbitrary finite field-dependent parameter. For instance, we choose an specific
 $\Theta[\phi]$ obtained from the following infinitesimal field-dependent parameter using relation (\ref{80})
\begin{eqnarray}
\Theta'[\phi]=  \int d^4 y \left[ \bar C^a B^a (B^2)^{-1} \left(\frac{\delta \xi}{2i }B^2 \right)\right].\label{th}
\end{eqnarray}
The finite field-dependent parameter is calculated by
\begin{eqnarray}
\Theta[\phi]=  \int d^4 y \left[ \bar C^a B^a (B^2)^{-1} \left(\exp\left\lbrace\frac{\delta \xi}{2i }B^2 \right\rbrace
-1\right)\right],
\end{eqnarray}
which coincides with the parameter given in \cite{lav}.

Exploiting the relations (\ref{jac}) and (\ref{th}), we 
compute the change in Jacobian as follows:
\begin{equation}
 \frac{d}{d\kappa}\ln J(\kappa)= - i\int  d^4x\left [ \frac{\delta \xi}{2  }B^2 \right], 
 \label{ja1}
\end{equation}
An ansatz for the (arbitrary) local functional $S_1[\phi]$  which appears in the 
expression (exponent) of the Jacobian (\ref{js}) as
\begin{equation}
S_1[\phi]= \int  d^4x\left [\chi(\kappa)  B^2 \right], 
\end{equation}
where $\chi(\kappa)$ is an arbitrary constant parameter 
constrained by 
\begin{equation}
\chi(\kappa=0)=0,\label{bo}
\end{equation}
so that the requirement (\ref{ini}) holds. Now, the change in $S_1$ with respect to $\kappa$
can easily be calculated by utilizing the relation (\ref{diff}) as follows:
\begin{eqnarray}
\frac{dS_1}{d\kappa}=\int d^4 x \left[ \frac{d\chi}{d\kappa}B^2\right].\label{s11}
\end{eqnarray}
The  existence of the   functional $S_1$ is valid when it  satisfies the essential requirement
given in (\ref{mcond}) along with   (\ref{ja1}) and (\ref{s11}). 
This leads to the following exactly solvable linear differential equation:
\begin{eqnarray}
\frac{d\chi}{d\kappa}+\frac{\delta\xi}{2 }=0.
\end{eqnarray}
The unique solution of the above equation satisfying the initial boundary condition
(\ref{ini}) is given by
\begin{equation}
\chi(\kappa)=-\frac{\delta\xi}{2 }\kappa.
\end{equation}
Having this specific value of $\chi$ the expression of $S_1$ at $\kappa =1$ reads
\begin{equation}
S_1[\phi]|_{\kappa=1}= -\int  d^4x\left [\frac{\delta\xi}{2 } B^2 \right].
\end{equation}
This implies that  the effective action
 within functional integral under the action of FFBRST transformation changes to
 \begin{eqnarray}
S_{\xi} +S_1[\phi]|_{\kappa=1}&=& \int d^4x \left [-\frac{1}{4}F^{a\mu\nu}F_{\mu\nu}^a +\frac{(\xi-\delta\xi)}{2}(B^a) ^2 + B^{a}
\partial^\mu {A^a_\mu }+ \bar{C}^a\partial^\mu
D^{ab}_\mu C^b\right ],\nonumber\\
&=:& S_{\xi -\delta\xi}, 
\end{eqnarray}
which is nothing but the effective YM action in $R_{\xi-\delta\xi}$ gauge.
However, for $\delta\xi =\xi$, the connection between Landau and $R_\xi$ can easily be  
established. 
Therefore,  FFBRST transformation with specific parameter 
mentioned in (\ref{th}) connects the $R_\xi$ and $R_{\xi-\delta\xi}$ gauges of the
YM theory. In summery we see that
\begin{eqnarray}
Z_\xi \left(\equiv\int {\cal D}\phi\ e^{iS_{\xi}}\right)\stackrel{FFBRST}{-----\longrightarrow}
Z_{\xi-\delta\xi} \left(\equiv\int {\cal D}\phi\ e^{iS_{\xi-\delta\xi}}\right).
\end{eqnarray}
Hence, we have obtained exactly same results of Ref. \cite{lav} using the formulation of Ref. \cite{bm}
which was developed long ago.
These results justifies the equivalence between the two
field-dependent BRST approaches, namely, the original FFBRST approach \cite{bm} and the recently
developed field-dependent BRST approach advocated in \cite{lav}.
\section{Conclusions}
The   finite field-dependent BRST (FFBRST) transformation was developed long ago
\cite{bm} and has been found great importance
in the different context of gauge field theories.  Recently, a new method  has been explored to study the field-dependent BRST transformation \cite{lav}. Although being presented differently we have found that these two approaches are exactly same. The novelty of these symmetry transformations are: they lead a non-trivial Jacobian for the functional measure under change of variables and capable of connecting the generating functional of
different effective theories.   In this paper, we have calculated the Jacobian 
for path integral measure using former FFBRST method which are found same to that of 
calculated by the later approach.
Further, we have shown that the FFBRST transformation amounts a precise change in the functional
integral of YM theory as shown in Ref. \cite{lav}.  Our study establishes the equivalence 
between the two approaches of FFBRST transformations.

\end{document}